\documentstyle[preprint,aps]{revtex}

\tightenlines 

\def\CQG{Class.\ Quant.\ Grav.\ }

\def\NPB{Nucl.\ Phys.\ }

\def\PLB{Phys.\ Lett.\  }

\def\PRD{Phys.\ Rev.\  }

\def\MPLA{Mod.\ Phys.\ Lett.\ }

\def\IJMPA{Int.\ Jou.\ Mod.\ Phys.\ }

\def\JHEP{J.\ High Energy Phys.\ }

\def\TMP{Theor.\ Math.\ Phys.\ }

\def\e{\epsilon}

\def\f{\phi}

\def\o{\omega}

\def\l{\lambda}

\def\la{\l^{2}}

\def\p{\pi}

\def\r{\rho}

\def\de{\partial}

\def\ds{ds^2=}\def\sg{\sqrt{-g}}

\def\sg{\sqrt{-g}}

\def\fo{\f_0}

\def\ord#1{O\left(#1\right)}

\def\r{\rho}

\def\ds{ds^2=}

\def\be{\begin{equation}}

\def\ee{\end{equation}}

\def\bea{\begin{eqnarray}}

\def\eea{\end{eqnarray}}

\def\bc{\begin{displaymath}}

\def\ec{\end{displaymath}}

\def\lb{\label}

\def\adsd{$\rm AdS_{2}$ }

\def\adsdp{$\rm AdS_{2}^{+}$ }

\def\adsdz{$\rm AdS_{2}^{0}$ }

\def\be{\begin{equation}}

\def\ee{\end{equation}}

\def\ba{\begin{eqnarray}}

\def\ea{\end{eqnarray}}

\def\two{^{(2)}}

\def\bz{\bar z}

\def\p{\partial}

\def\pt{\partial_t}

\def\px{\partial_x}

\def\Lo{l^{(2)}_{0}}

\begin{document}
\preprint{\vbox{\noindent
\null\hfill INFNCA-TH0009}}

%

%


\draft

\vskip 3truecm

\title{Two-dimensional black holes as open strings: A new realization  of the 
ADS/CFT correspondence\thanks{This work is supported in part by funds provided
by the U.S.\ Department of Energy (D.O.E.) under cooperative research agreement
DE-FC02-94ER40818 and by a FCT grant, contract number BPD/20166/99.}}

\author{Mariano Cadoni$^{a}$\footnote{E--Mail: cadoni@ca.infn.it},
Marco Cavagli\`a$^{b,c}$\footnote{Post office address:
Massachusetts Institute of Technology,
Marco Cavagli\`a 6-408, 77 Massachusetts Avenue,
Cambridge MA 02139-4307, USA. E-mail: cavaglia@mitlns.mit.edu}}
\address{$^a$ Dipartimento di Fisica, Universit\`a di Cagliari,\\
Cittadella
Universitaria 09042, Monserrato, Italy and INFN, Sezione di Cagliari\\
$^b$ Departamento de F{\'\i}sica, Universidade da Beira Interior\\ 
R.\ Marqu\^es d'\'Avila e Bolama, 6200 Covilh\~a, Portugal\\
$^c$ Center for Theoretical Physics,
Laboratory for Nuclear Science and Department of Physics,
Massachusetts Institute of Technology,
77 Massachusetts Avenue, Cambridge MA 02139, USA and 
INFN, Sede di Presidenza, Roma, Italia.}
\vskip 3truecm




\maketitle

\begin{abstract}
We show that weak-coupled two-dimensional dilaton gravity on 
Anti-de Sitter space can be described by the dynamics of an open string. 
Neumann and Dirichlet boundary conditions for
the string lead to two different realizations of the Anti-de
Sitter/Conformal Field Theory correspondence. In particular, in the
Dirichlet case the thermodynamical entropy of two-dimensional black
holes can be exactly reproduced by counting the string states.
\end{abstract}





The realization of the holographic principle in two spacetime dimensions
is a subject that has recently attracted much attention in the
literature, where it has been mainly investigated in the context of the
Anti-de  Sitter/Conformal Field Theory   (AdS$_{d}$/CFT$_{d-1}$)
correspondence \cite{Wm}.  For $d=2$ it states that gravity on \adsd is
dual to a one-dimensional conformal field theory living on the boundary
of \adsd.

In spite of the efforts that have been devoted to clarify the
AdS$_{2}$/CFT$_{1}$ duality \cite{g1,g2,CM,CM1}, the latter remains
puzzling and mysterious. Since the conformal symmetry involved in the
duality is infinite dimensional, the dynamics is expected to be highly
constrained. However, the realization of the symmetry in terms of
boundary states is far from trivial \cite{CM,CM1}. This difficulty
seems related to the topology of the boundary of \adsd, which is
one-dimensional and disconnected.

The lack in understanding of the AdS$_{2}$/CFT$_{1}$ duality has prevented real
progress in what is considered its main application: the study of
two-dimensional (2D) gravity structures (e.g black holes) \cite{KS} using
conformal field theory techniques. This application is of fundamental relevance
for black holes physics because it can be used to give statistical meaning to
the entropy of both 2D black holes and higher dimensional black holes that
reduce to 2D models upon compactification. Attempts to calculate the
statistical entropy of 2D \adsd black holes met only partial success
\cite{CM,CM1}. A mismatch of a factor $\sqrt 2$ between the thermodynamical and
statistical entropy was found.

In this letter we clarify the meaning of the AdS/CFT  correspondence 
in two
dimensions by showing that it can be realized in two different ways both
steming from a more fundamental AdS$_{2}$/CFT$_{2}$ correspondence.
Using the nonlinear sigma model formulation of 2D dilaton gravity
\cite{cav1} we show that weak-coupled dilaton gravity on \adsd can be 
described by the dynamics of an open string. Using Neumann boundary conditions 
we retrieve the
AdS$_{2}$/CFT$_{1}$ correspondence that has been analyzed in Ref.\
\cite{CM,CM1}. Dirichlet boundary conditions lead to a new
realization of the AdS/CFT correspondence. In this case the properties of 2D
black holes have a direct interpretation in terms of string dynamics. In
particular, the entropy of the black hole can be exactly computed in
terms of the degeneracy of the open string spectrum.

The simplest 2D gravity model admitting \adsd as solution is the 
Jackiw-Teitelboim (JT) model \cite{JT}
\be\lb{e1}
A={1\over2}\int \sg \, d^2x\, \f\left(R+2\la\right)\,.    
\ee
The classical solutions of the model,
\be\lb{e2}
\ds-\left(\l^2r^2- {2m\over \l \fo}\right)dt^2+\left(\l^2r^2-
{2 m\over \l \fo}\right)^{-1}dr^2\,,
\quad \f=\fo \l r\,,\quad m\ge 0\,,
\ee
can be interpreted as \adsd-black holes \cite{CM2}. The black hole mass
$m$ appearing in Eq. (\ref{e2}) is defined by the mass functional
\be\lb{e3}
M=  \la \f^{2}- \de_{\r}\f  \de^{\r}\f.  
\ee
On-shell $M$ is constant \cite{mass} and equal to $2\fo \l m$.

The 2D gravity model (\ref{e1}) is pure gauge, i.e., it has no
physical local degrees of freedom. Moreover, solutions (\ref{e2}) with
different values of $m$ represent different, locally equivalent,
parametrization of \adsd. However, the presence of the scalar $\phi$
makes them globally nonequivalent \cite{CM2}. Following the notation
of Ref.\ \cite{CM2} we will denote with \adsdp and \adsdz the black
hole solutions ($m>0$) and the ground state ($m=0$), respectively.

The link between 2D AdS-gravity and  CFT can be  established using the
asymptotic symmetries of \adsd. It has been shown in Refs.\
\cite{CM,CM1} that the asymptotic symmetries of \adsd are generated by a
Virasoro algebra and that the deformations of the timelike boundary of
\adsd give a realization of the conformal symmetry. In Refs.\
\cite{CM,CM1} the $(r,t)$ coordinates of Eq. (\ref{e2}) have been used
to discuss the asymptotic symmetries of \adsd, yet for our  purposes it
is convenient to use light-cone coordinates $(u,v)$. In the
$(u,v)$-frame the \adsdz solution is $g_{uv}=2/\l^2 (u+v)^2$,
$g_{uu}=g_{vv}=0$, $\phi=-\phi_0 / \lambda(u+v)$, so the boundary
conditions \cite{CM} to be imposed on the metric and on the dilaton are
$g_{uv}=2/\l^2 (u+v)^2+\ord1$, $g_{uu}=\ord 1$, $g_{vv}=\ord1$,  and
$\phi=\ord {(u+v)^{-1}}$, respectively. The metric and the dilaton have
the asymptotic, $u\to -v$, form
\bea\lb{e4}
g_{uu}&=&U_0+\ldots +U_n(u+v)^{n}+\ldots\,,\nonumber\\
g_{uv}&=&{2\over \lambda^2(u+v)^2}+Y_0+\ldots +Y_{n}(u+v)^{n}+
\ldots\,,\nonumber\\
g_{vv}&=&V_0+\ldots +V_n(u+v)^{n}+\ldots\,,\nonumber\\
\phi&=&-\phi_0\left[{\omega_{-1}\over \lambda
(u+v)}+\omega_{1}\lambda(u+v)+\ldots+ 
\o_{n}\lambda^{n}(u+v)^{n}+\ldots\right]\,,
\eea
where the coefficients $U_{k},U_{k}, \omega_{k}$ are functions of
$u-v$  only.  The transformations generated by the asymptotic symmetry
group leave unchanged the leading  terms in Eq. (\ref{e4}) and act on
the remaining functions  $U_{k},V_{k},Y_{k}$ and  $\omega_{k}$. These
can be thought as characterizing the deformations of the $u=-v$
boundary of \adsd.

The asymptotic symmetry group is generated  by the Killing vectors
\bea\lb{e5}
\chi^u &=&{1\over 2}\left[ \epsilon+\epsilon'(u+v)+
{1\over 2}\epsilon''(u+v)^2 \right]+ \alpha^u\,,\nonumber\\
\chi^v&=& {1\over 2}\left[-\epsilon+\epsilon'(u+v)- 
{1\over 2}\epsilon''(u+v)^2 \right]+\alpha^v\,,
\eea
where $\epsilon\equiv\epsilon(u-v)$ is an arbitrary function,
$\alpha^{u,v}= \sum_{k=3}^{\infty} \alpha^{u,v}_{k}(u-v) (u+v)^{k}$,
and $'={d/ d(u-v)}$. $\alpha^{u,v}$ represent ``pure gauge''
diffeomorphisms of the 2D  gravity theory that fall off rapidly as
$u\to -v$. The Killing vectors (\ref{e5}) define a conformal group
generated  by a Virasoro algebra and the boundary fields
$\Theta_{k}=(U_{k},V_{k},Y_{k},\omega_{k})$ span a representation  of
this symmetry. Their transformation law has the form
\be\lb{e6}
\delta_\e \Theta_{k}= \epsilon \Theta_{k}'+(h+k)\epsilon'\Theta_{k}+\ldots\,,
\ee
where $h=2$ for the fields $U,V,Y$ and $h=0$ for the fields $\omega$,
and dots denote terms that depend on higher derivatives of $\e$ and on
the ``pure gauge'' diffeomorphisms. It is important to notice that
although ``pure gauge'' transformations affect the boundary fields, the
charges that are associated with the asymptotic symmetry are invariant
\cite{CM1} under pure gauge transformations. The mass functional $M$ can
be likewise expanded near the $u=-v$ boundary
\be\lb{e7}
M=\sum_{k=0}^{\infty} M_{k}(u-v) (u+v)^{k}\,.
\ee
Using Eqs.\ (\ref{e3}) and (\ref{e4}) the $M_{k}$ can be expressed in
terms of the boundary fields. They follow the transformation law
(\ref{e6}) with $h=0$.

The action (\ref{e1}) can be cast in the form of a 2D
conformal nonlinear sigma model \cite{cav1} with Lagrangian
\be\lb{e8}
{\cal L}=\sqrt{-g}\partial_\mu M\partial^\mu\psi\cdot{1\over
1-4\lambda^2\psi^2\,M}\,,
\ee
where $\phi=(-2\lambda^2\psi)^{-1}$ and we have neglected irrelevant
surface terms. The Lagrangian (\ref{e8}) can be expanded around $\psi=0$
\be\lb{e9}
{\cal L}=\sqrt{-g}\sum_{k=0}^\infty \partial_\mu M_k\partial^\mu\psi_k\,,
\ee 
where $M_k=M^{k+1}/(k+1)$, $\psi_k=(2\lambda\psi)^{2k+1}/2\lambda(2k+1)$.
Equation (\ref{e9}) is both a perturbative expansion in terms of the
(coordinate-dependent) gravitational coupling of the model (\ref{e1})
($\phi^{-1}$) and an expansion near the boundary of \adsd. Each term in
Eq.\ (\ref{e9}) has the form of a free-field conformal theory and
transforms according to Eq.\ (\ref{e6}) with $h=2$, although the sum
(full theory) does not. In the weak-coupling regime, $\psi<<1$, the
theory can be treated perturbatively. In particular, if we restrict
ourselves to the first order in the perturbative expansion the theory
reduces to a free CFT and the usual machinery of CFTs can be applied. In
this sense we speak of ``duality" between (weak-coupled) \adsd\ gravity
and CFT.  Let us stress that the identification of a weak-coupling
regime,  where a consistent perturbative expansion can be constructed, is
a fundamental  feature of the sigma model representation that does not
have counterpart in the original formulation based on Eq.\ (\ref{e1}) and
is essential for the identification of the fundamental microscopic
degrees of freedom of the theory.  In this paper  we shall focus        
attention on the first term in the perturbative expansion (\ref{e9})
leaving the discussion of higher terms to further investigations.  This
amounts to neglecting, in first approximation, higher order perturbative
corrections generated by the running gravitational coupling.  Note that
the latter becomes strong near $r=0$,  i.e., precisely in the ``opposite"
region of $AdS_2$  around which we are expanding. In this approach higher
order corrections are described by interacting terms for the free CFT.

It is convenient to define the new fields
\be\lb{e10}
\begin{array}{lllll}
\sqrt{\pi\alpha'}M_0&=&\displaystyle{1\over 2}(X^1+iX^2)&=&\displaystyle
{1\over 2}(X^0+X^1)\,,\\
\sqrt{\pi\alpha'}\psi_0&=&\displaystyle{1\over 2}(X^1-iX^2)&=&\displaystyle
{1\over
2}(X^1-X^0)\,,
\end{array}
\ee
where $\alpha'$ is a constant with dimension of $lenght^2$. Using complex
coordinates $z\equiv u=(t+x)/2= (\sigma^1+i\sigma^2)/2$ and $\bar z\equiv
v=(x-t)/2=(\sigma^1-i\sigma^2)/2$, the leading term in the expansion (\ref{e9})
can be cast in the usual bosonic string form. [See Ref.\ \cite{pol} for
notations]. Since \adsd has a timelike boundary at $x=0$, we are dealing with
open strings and the expansion (\ref{e9}) defines a  AdS$_{2}$/CFT$_{2}$
correspondence between open string theory and dilaton gravity on AdS$_{2}$.
Boundary conditions are restricted to Dirichlet ($X^{\mu}(x=0)=const$) or
Neumann ($n^{a}\partial_{a} X^{\mu}(x=0)=0$) type, respectively.  Mixed
boundary conditions are not allowed because from the boundary  expansion
(\ref{e4}) for the field $\phi$ it follows $\partial_{t}X^{0}(x=0)=
\partial_{t}X^{1}(x=0)$. The choice of boundary conditions determines the
realization of the AdS/CFT correspondence. The AdS$_{2}$/CFT$_{1}$
correspondence that has been proposed in Ref.\cite{CM} is obtained by imposing
Neumann boundary conditions, which allow for excitations on the boundary. In
this case we have $X^{\mu}(x=0)=F(t)=M_{0}$ and the conformal symmetry can be
realized on the boundary by the charges that are  associated with the
asymptotic symmetries of \adsd. Dirichlet boundary conditions break translation
invariance in the $x$ direction and no dynamical degrees of freedom are allowed
on the boundary, the string endpoint being fixed. In this case we are naturally
lead to a new realization of the AdS/CFT correspondence. It is shown below that
the correspondence is  realized in terms of pure deformations of the boundary
of AdS$_{2}$.

In addition to the timelike boundary at $x=0$, AdS$_{2}^{0}$ has an inner null
boundary. However, the presence of the latter does not influence the dynamics
of the open string. Writing the solution, Eq.\ (\ref {e2}), in the conformal
coordinate frame $(t,x)$, one finds that AdS$_{2}^{0}$  is conformal to
Minkowski space and that the presence of the  dilaton requires
\be\lb{boun}
-\infty<t<\infty\,,\qquad 0\le x<\infty\,.
\ee
In this coordinate frame the inner null boundary is located at $x=\infty$.
Hence, because of conformal invariance, open strings on \adsdz\ are 
equivalent to open
strings on the region of the  Minkowski spacetime defined by Eq.\
(\ref{boun}).

The AdS$_{2}$/CFT$_{2}$ correspondence is expressed in a exact form by
putting in a one-to-one correspondence the symmetries and local
degrees of freedom of the open string and the asymptotic symmetries and
excitations of \adsd. The conformal symmetry in two spacetime
dimensions  is generated by the  Killing vectors,
$\chi=\chi(z)\partial+\bar\chi(\bar z)\bar\partial$. A generic
CFT$_{2}$ field $\psi(z,\bar z)$ of weights $(h,\bar h)$ transforms as
\be\lb{e11}
\delta_{\chi,\bar\chi}\psi= (\chi\partial+h\partial\chi)\psi+
(\bar\chi\bar\partial+\bar h\bar\partial\bar\chi)\psi\,.
\ee  
Dirichlet boundary conditions require that $\chi$ and $\bar\chi$ are 
related by the condition
\be\lb{e12}
\chi(z)+\bar\chi(\bar z)=0\,.
\ee
This equation implies that the conformal symmetry is generated by a 
single copy of the Virasoro algebra. By expanding the Killing vectors on
the boundary we obtain Eq.\ (\ref{e5})  with $\chi((z-\bar z)/2)=-\bar
\chi(-(z-\bar z)/2)=\epsilon(u-v)/2$, where the pure gauge
diffeomorphisms have been fixed as
$\alpha^{u}_{k}=(-1)^{k+1}\alpha^{v}_{k}=(1/2k!){d^{k}\epsilon/
d(u-v)^{k}}$. Thus, by fixing the pure gauge diffeomorphisms
appropriately, the symmetry group of the Dirichlet open string and the
asymptotic symmetry group of AdS$_{2}$ coincide. Each \adsd field living
near $x=0$ can be interpreted as the coefficient of the expansion of the
CFT$_2$ field around the boundary with given weight $h+\bar h$ and pole
of order $p$. Moreover, the above  correspondence allows to determine
from CFT$_2$ the Virasoro generators of the asymptotic symmetry group of
AdS$_{2}$. Using Eq.\ (\ref{e12}) and expressing the CFT$_{2}$ Virasoro
generators $L^{CFT}_m=z^{-m+1}\partial$ and $\tilde
L^{CFT}_{m}=\bz^{-m+1}\bar\partial$ as functions of the $x,t$
coordinates we find
\be\lb{e13}
L^{AdS}_{m}=2^{m-1}\left\{\left[(t+x)^{-m+1}+(t-x)^{-m+1}\right]\pt+
\left[(t+x)^{-m+1}-(t-x)^{-m+1}\right]\px\right\}\,.
\ee
The AdS Virasoro generators (\ref{e13}) are valid both on the boundary
and outside the boundary, where they generate the full symmetry group of
the open string with Dirichlet boundary conditions. Equation (\ref{e13})
leads to the asymptotic AdS Killing vectors (\ref{e5}) with fixed gauge
diffeomorphisms. By fixing the pure gauge diffeomorphisms of the AdS
asymptotic symmetries we can reconstruct the full symmetry group of the
Dirichlet open string.  According to this picture the Virasoro generators
$L^{AdS}_{m}$ cannot be interpreted as generating the   symmetries of a
1D conformal field theory living on the boundary of \adsd, the latter
being frozen by the Dirichlet boundary conditions.

The AdS$_{2}$/CFT$_{2}$ correspondence can also be realized using local
oscillator degrees of freedom. Let us expand the string field in normal
modes
\be\lb{e14}
X^{\mu}=x^{\mu} -ip^{\mu}\log|z|^{2}+i \left(\alpha'\over 2\right)^{1/2}
\sum_{m=-\infty}^{\infty}{1\over m}\left(\alpha^{\mu}_{m}z^{-m}+\tilde
\alpha^{\mu}_{m}\bar z^{-m}\right)\,.
\ee
Comparing Eq.\ (\ref{e14}) to the asymptotic expansions of $M_0$ and
$\psi_0$
\be\lb{e15}
M_0=\sum_{k=0}^{\infty}\sum_{m=-\infty}^{\infty}
M_{km}x^{k}t^{m}\,,\qquad
\psi_0= \sum_{k=1}^{\infty}\sum_{m=-\infty}^{\infty}
\Psi_{km}x^{k}t^{m}\,,
\ee
we find [we assume $t>0$ for simplicity]
\be\lb{e16}
\alpha^\mu_m=(-1)^{m+1}\tilde\alpha^\mu_m=i\sqrt{\pi}2^{-1/2-m}
\left[M_{1,-1-m}\mp\Psi_{1,-1-m}\right]\,,
\ee
where we have imposed the Dirichlet boundary conditions that imply
$p^{\mu}=0$, $M_{00}=const$, and $M_{0 m}=0$ for $m\neq 0$. The
asymptotic excitations of the gravity theory are in a one-to-one
correspondence with the open string modes. Moreover, the lower terms
in the asymptotic expansion are sufficient to determine the whole
CFT$_{2}$ theory. The fields $M_1$, $\Psi_1$ and $M_0$ are invariant
under pure gauge bulk transformations and transform conformally with weight
$h=1$ ($M_1$ and $\Psi_1$) and $h=0$ ($M_0$). Therefore,  Eq.\
(\ref{e16}) provides a realization of the
AdS$_{2}$/CFT$_{2}$ correspondence: Asymptotic 2D gravity modes around
the boundary that describe boundary deformations determine completely
CFT$_{2}$ (the open string theory) and vice versa. Finally, using Eq.\
(\ref{e16}) the CFT$_{2}$ Virasoro generators can be expressed in
terms of the asymptotic modes
\be\lb{e17}
L^{CFT}_m={1\over2}\sum_{n=-\infty}^\infty \alpha^\mu_{m-n}\alpha_{\mu n}=
 -\pi 2^{-m}\sum_{n=-\infty}^\infty M_{1,-1-n}\Psi_{1,-1-m+n}\,.
\ee
By imposing Neumann boundary conditions on the open string, the string
modes are determined by the gravitational modes $M_{0,m}$. In
this case the Virasoro generators (\ref{e17}) and the  CFT$_{2}$
action are zero at any order in the expansion and the conformal
symmetry cannot be realized in terms of local string oscillators.
Rather, we are dealing with a topological theory which has no physical
local degrees of freedom and the conformal symmetry is realized by the
charges associated with the asymptotic symmetries \cite{CM}.

The AdS$_{2}$/CFT$_{2}$ correspondence leads to a natural interpretation
of the Hawking evaporation of the \adsd black hole \cite{CM2}. From Eq.\
(\ref{e13}) it follows that the invariant $SL(2,R)$ algebra is generated
by
\be\lb{e18}
L^{AdS}_{0}=t\pt+x\px\,,\qquad
L^{AdS}_{1}=2\pt\,,\qquad
L^{AdS}_{-1}={1\over 2}(t^2+x^2)\pt+xt\px\,.
\ee
In the representation above $L^{AdS}_{0}$ does not generate translations
in $t$ but dilatations. $L^{AdS}_{0}$ generates time translations
in the $T,X$ coordinates
\be\lb{e18a}
\l t=e^{\l T}\cosh(\l X)\,,\qquad\l x=e^{\l T}\sinh(\l X)\,.
\ee
Using Eq.\ (\ref{e18a}) the metric of the \adsdz ground state is
\be\lb{e19}
ds^2={1\over\sinh^2(\l X)}(-dT^2+dX^2)\,.
\ee
Equation (\ref{e19}) describes a 2D black hole \cite{CM2}. Hence, the
Hawking  evaporation process can be explained in the CFT$_{2}$ context
using the same arguments of Ref.\ \cite{CM2}. Positive frequency modes
of a quantum field with respect to Killing vector $\partial_{t}$ are
not positive frequency modes with respect to Killing vector
$\partial_{T}$, i.e., the vacuum state for an observer in the $(X,T)$
reference frame appears filled with thermal radiation to an observer
in the $(x,t)$  frame. The value of the Hawking flux has been
calculated in Ref. \cite{CM2}.

The correspondence between the open string with Dirichlet boundary conditions 
and the 2D \adsd black hole can be used to calculate the statistical
entropy  of the latter. Since local oscillators of the Dirichlet string
are in one-to-one correspondence with excitations of \adsdz we can count
black hole states by counting states of CFT$_{2}$. To this purpose, we
calculate the central charge $c$ associated with the central extension of
the Virasoro algebra generated by $L^{CFT}_{m}$. Keeping in mind the
interpretation of $c$ as a Casimir energy (see for instance Ref.\
\cite{pol}), the transformation law of the stress-energy tensor under the
change of coordinates (\ref{e18a}) ($w=T+X$) is
\be
 T\two_{ww}=(\p_{w}z)^{2}T\two_{zz}- {c\over 12} \{w,z\}(\p_{w}z)^{2}\,.
\ee
The vacuum energy is shifted by $\Lo\to\Lo - {c\over 24}$, where $\Lo$ is
the eigenvalue of $ L^{CFT}_{0}$ which is associated to the vacuum.
This shift corresponds to a Casimir energy $E=- {c\over 24}\l$.

The coordinate transformation (\ref{e18a}) maps the \adsdz ground state
solution of the 2D dilaton gravity theory into the \adsdp black hole 
solution (\ref{e19}) with mass $m={\fo\over 2} \l$ (see Ref.\
\cite{CM2}). Because of the  correspondence between the gravitational
theory and the Dirichlet string we can interpret the previous map as
the gravity theory counterpart of the shift of $ L^{CFT}_{0}$ in CFT$_2$ 
and equate the Casimir energy $E$ with $m$. There is a subtlety
concerning the sign to be used in the equation. The coordinate
transformation  (\ref{e18a}) is analogous to the coordinate
transformation that maps  the Rindler spacetime into the Minkowski
spacetime, i.e., it maps {\it observers}. So an observer in the \adsdp
vacuum sees the \adsdz  vacuum as filled with thermal radiation with
{\it negative} flux \cite{CM2}. Since the Casimir energy $E$ is the
energy of the $z$-vacuum as seen in the $w$-frame, we must use the
equation $E=-m$ which leads to $c=12\fo$. Finally, the eigenvalue of $
L^{CFT}_{0}$ can be expressed in terms of the black hole mass. Using the
Cardy  formula \cite{ca} the statistical black hole entropy is
\be
S=2\pi \sqrt { c L^{CFT}_{0}  \over 6}= 4\pi \sqrt{ \fo m\over 2\l}\,, 
\ee
in complete agreement with the thermodynamical result.

In this letter we have proved that the correspondence  between 2D gravity and
open strings allows for two distinct realization of the AdS/CFT
correspondence. The first realization, which is obtained by imposing
Neumann boundary conditions to the open string, implies the existence
of a genuine one-dimensional CFT living on the boundary of \adsd. This
realization is, however, problematic from different points of view
\cite{CM,CM1}. 
The realization which is obtained by imposing Dirichlet boundary
conditions supports the viewpoint of Ref.\  \cite{caccia}, where, by
quite a different argument,  the authors conclude that the correspondence
should be realized as \adsd/CFT$_{2}$, rather than \adsd/CFT$_1$.

\end{document}